\documentclass[aps,prl,twocolumn,superscriptaddress,showpacs,amsmath,amssymb]{revtex4}

\usepackage[dvipdfmx]{graphicx}
\usepackage{graphicx}

\begin{document}


\title{Real Space Imaging of Spin Polarons in Zn Doped SrCu$_2$(BO$_3$)$_2$}


\author{M. Yoshida}
\affiliation{Institute for Solid State Physics, University of Tokyo, Kashiwa, Chiba 277-8581, Japan}
\author{H. Kobayashi}
\affiliation{Institute for Solid State Physics, University of Tokyo, Kashiwa, Chiba 277-8581, Japan}
\author{I. Yamauchi}
\affiliation{Institute for Solid State Physics, University of Tokyo, Kashiwa, Chiba 277-8581, Japan}
\author{M. Takigawa}
\affiliation{Institute for Solid State Physics, University of Tokyo, Kashiwa, Chiba 277-8581, Japan}
\author{S. Capponi}
\affiliation{Laboratoire de Physique Th\'{e}orique, Universit\'{e} de Toulouse and CNRS, UPS (IRSAMC), F-31062 Toulouse, France}
\author{D. Poilblanc}
\affiliation{Laboratoire de Physique Th\'{e}orique, Universit\'{e} de Toulouse and CNRS, UPS (IRSAMC), F-31062 Toulouse, France}
\author{F. Mila}
\affiliation{Institut de Th\'{e}orie des Ph\'{e}nom\`{e}nes Physiques, \'{E}cole Polytechnique F\'{e}d\'{e}rale de Lausanne, CH-1015 Lausanne, Switzerland}
\author{K. Kudo}
\affiliation{Department of Physics, Okayama University, Okayama 700-8530, Japan}
\author{Y. Koike}
\affiliation{Department of Applied Physics, Tohoku University, Sendai 980-8579, Japan}
\author{N. Kobayashi}
\affiliation{Institute for Materials Research, Tohoku University, Sendai 980-8577, Japan}


\date{\today}

\begin{abstract}
We report on the real space profile of spin polarons in the quasi two-dimensional frustrated 
dimer spin system SrCu$_2$(BO$_3$)$_2$ doped with 0.16\% of Zn. The $^{11}$B nuclear magnetic resonance 
spectrum exhibits 15 additional boron sites near non-magnetic Zn impurities. With the help of exact diagonalizations
of finite clusters, we have deduced from the boron spectrum the distribution of local magnetizations at the
Cu sites with fine spatial resolution, providing direct evidence for an extended spin polaron. The results are confronted with 
those of other experiments performed on doped and undoped samples of SrCu$_2$(BO$_3$)$_2$.
\end{abstract}


\maketitle



Impurities and defects in strongly correlated quantum systems often produce significant effects over an 
extended spatial region, which can be studied by local probes such as nuclear or 
electron magnetic resonance (NMR or ESR) \cite{Alloul}. The best example is the edge states in
Heisenberg spin chains. The spin 1/2 edge state in spin 1 Haldane chains is a direct consequence of the 
valence-bond-solid ground state of the pure system. The ESR experiments have played vital roles in identifying 
the edge spins \cite{Hagiwara,Glarum} and their interactions \cite{Yoshida}. The edge states are not 
localized at a single site but associated with local staggered magnetization due to the
antiferromagnetic interaction of the bulk, and the spatial extent of such a polaronic structure is given by 
the correlation length of the bulk. The real space profile of spin polarons has been actually observed 
by NMR experiments in both spin 1 \cite{Tedoldi, Das} and spin 1/2 \cite{Takigawa} Heisenberg chains, from which 
the temperature dependence of the correlation length was deduced. 

Although there have been less studies on two-dimensional (2D) systems, an interesting example is the 
frustrated 2D dimer spin system SrCu$_2$(BO$_3$)$_2$ with a small concentration of 
Cu$^{2+}$ ions (spin 1/2) replaced by non-magnetic Zn or Mg \cite{Kudo,Haravifard,Shawish,Capponi}. 
The magnetic layers contain orthogonal arrays of Cu dimers described 
by the Shastry-Sutherland lattice \cite{Shastry}
\begin{equation}
H = J\sum_{n.n.}{\bf S}_i\cdot {\bf S}_j + J^{\prime}\sum_{n.n.n.}{\bf S}_i\cdot {\bf S}_j,
\label{SS}
\end{equation}
where $J$ ($J^{\prime}$ ) is the intradimer (interdimer) Heisenberg exchange interaction.
The ground state of SrCu$_2$(BO$_3$)$_2$ at zero magnetic field is the dimer singlet state \cite{Kageyama1,Kodama1}, 
which is known to be the exact ground state of Eq.~(\ref{SS}) for $\alpha=J^{\prime}/J$ not too large \cite{Shastry,Miyahara1},
less than $\alpha_c \simeq 0.675$ \cite{Koga,Corboz1}. SrCu$_2$(BO$_3$)$_2$
exhibits a number of fascinating properties, most notably a unique sequence of quantized magnetization 
plateaus in magnetic fields \cite{Onizuka,Kodama2,Sebastian,Jaime,Takigawa2,Matsuda} which have been a subject of intense research in
the last decade \cite{Miyahara3,Takigawa4}. 

A nonmagnetic impurity creates an unpaired Cu$^{2+}$ site in the dimer singlet state, producing a free 
spin-1/2. The structure factor of this spin-1/2 measured by inelastic neutron scattering
experiments \cite{Haravifard} points to an extended object.  Theories have confirmed this picture and
moreover predicted the formation of a spin polaron extending over several sites around the impurity \cite{Shawish,Capponi},
clearly calling for further precise experimental information.

In this letter, we report the observation 
of such a spin polaron in real space by $^{11}$B NMR experiments on Zn doped SrCu$_2$(BO$_3$)$_2$ performed
in a sufficiently high magnetic field to saturate unpaired spins. With the help of exact diagonalization results,
a nearly complete assignment of the 15 additional boron sites has been achieved, leading to the determination of 
the microscopic structure of a localized spin polaron with unprecedented accuracy. 

\begin{figure*}[t]
\includegraphics[width=0.9\linewidth]{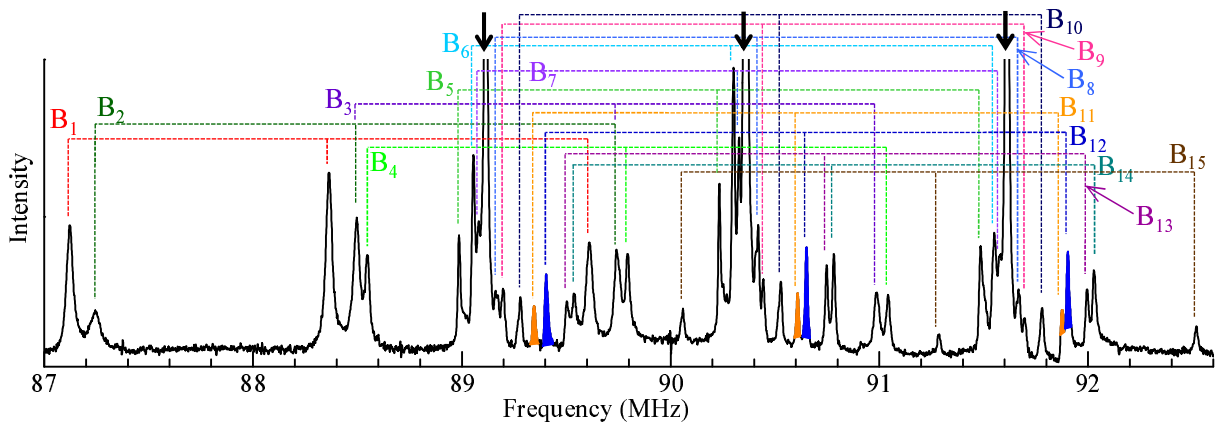}
\caption{(color online) $^{11}$B NMR spectrum at $T$ = 1.6 K and $B$ = 6.615 T. The black arrows mark the position 
of the reference line (zero internal field). 15 additional lines named B$_1$ to B$_{15}$ have been resolved. 
Note that B$_1$ is about twice as intense as B$_2$, and that
B$_{12}$ (shaded in blue) is also about twice as intense as B$_{11}$ (shaded in orange).}
\label{spectrum} 
\end{figure*}

Single crystals of SrCu$_{2-x}$Zn$_x$(BO$_3$)$_2$ were grown by the traveling-solvent 
floating-zone method \cite{Kageyama2,Kudo2}. Two crystals were used, $x$ = 0.0174 and 0.0032 
as determined by the inductively coupled plasma atomic emission spectrometry. 
The presence of free spins at low temperatures was confirmed by magnetization measurements 
(see Supplementary Material (Suppl. Mat.) A \cite{Supplement}). 
The crystals were cut into a rod 
($1 \times  1 \times  5 $ mm$^{3}$) for NMR measurements, which were performed in a 
magnetic field $B$ of 6.615~T precisely along the $c$ axis (within $\sim$0.2 degree). 

The NMR spectra were obtained by summing the Fourier transform of the spin-echo signal 
obtained at equally spaced rf-frequencies. 
Figure~\ref{spectrum} shows the $^{11}$B NMR spectrum for $x$ = 0.0032 
(0.16\% of Zn) at 1.6 K. The Zeeman energy for the magnetic field of 6.615~T is much smaller than
the zero-field energy gap for the triplet excitation in the bulk ($\Delta$ = 35~K) but large enough to
completely polarize the impurity induced free spins 
(see Suppl. Mat. A and B \cite{Supplement}). 
To understand the $^{11}$B NMR spectra, we first recall that one boron site generates three NMR 
lines at the frequencies $\nu _r = \gamma (B + B_{\mathrm{int}}) + k\nu _Q$, ($k = -1$, 0, 1), 
where $\nu_Q$ is the quadrupole splitting along the $c$ axis, 
$\gamma$ = 13.66 MHz/T is the nuclear gyromagnetic ratio, and $B_{\mathrm{int}}$ is the
internal magnetic field produced by nearby Cu spins. 
Since the Zn concentration is extremely dilute, most of Cu spins form singlet dimers generating
$B_{\mathrm{int}} \sim 0$ at the majority of B sites. 
The NMR lines from these B sites (shown by black arrows) are very intense, far exceeding the range of display in Fig.~\ref{spectrum}. 

In addition to this reference line, we have been able to identify 15 weaker lines with non-zero 
$B_{\mathrm{int}}$ (B$_{1}$ - B$_{15}$, the thin lines in Fig.~\ref{spectrum}) and to determine the 
values of $B_{\mathrm{int}}$ and $\nu _Q$ for each of them.  The sample with $x$ = 0.0174 gives a 
nearly identical NMR spectrum (see Suppl. Mat. C \cite{Supplement}), ensuring no interference between impurities. 

\begin{figure}[t]
\includegraphics[width=1.\linewidth]{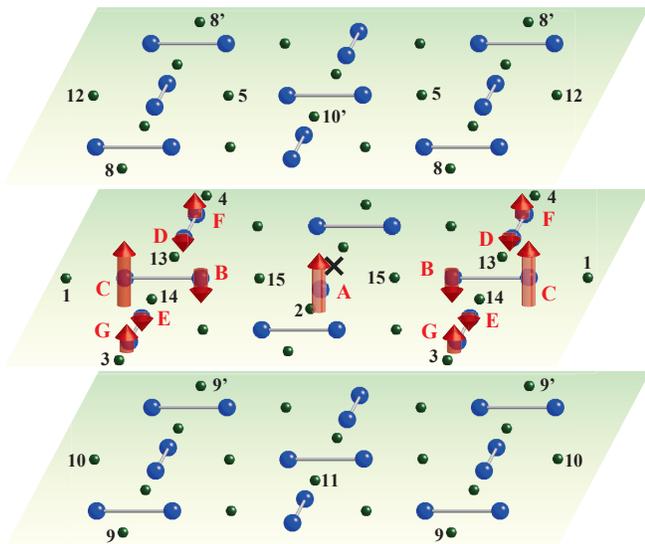}
\caption{(color online) Real space sketch of the spin polaron formed around a Zn impurity (cross). The up (down) arrows 
on the Cu sites represent the spin moments parallel (antiparallel) to the external field. The length of the arrow is proportional 
to $|\langle S_z \rangle |$ as calculated on a 36-site cluster with $J'/J=0.67$. 
The numbers show the assignment of the B sites to the NMR lines of Fig.~\ref{spectrum} deduced from the analysis
described in the text.  Primed numbers have been used when different sites are assigned to the same line.
The other B sites have very small internal field. }
\label{polaron}
\end{figure}

\begin{figure}[t]\
\includegraphics[width=0.8\linewidth]{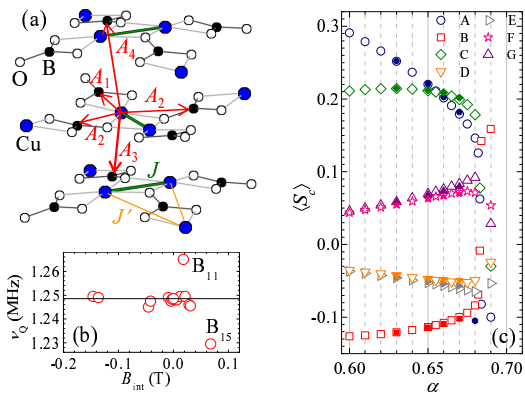}
\caption{(color online) (a) Main hyperfine couplings between a Cu spin and the B nuclei in the same layer 
($A_{1}$ and $A_{2}$) or in the neighboring layers ($A_{3}$ or $A_{4}$). (b) Quadrupolar splitting $\nu _Q$ for B sites 
near a Zn impurity. (c) Dependence on $\alpha=J'/J$ of the local magnetization calculated with exact diagonalizations 
on a 32-site cluster (open symbols) and 36-site cluster (solid symbols).} 
\label{ED} 
\end{figure}

As we shall demonstrate, it is possible to assign most of the lines to specific boron sites, and to deduce the 
polarization of the Cu sites around the impurity as shown in Fig.~\ref{polaron}.
To perform this line assignment, it is useful to know {\it a priori} the local magnetization expected
in the neighborhood of a Zn impurity. We have thus performed exact diagonalizations (ED) calculation for 
finite-size clusters of the 2D Shastry-Sutherland lattice with 32 sites (31 spins and one vacancy) and 36 sites 
(35 spins and one vacancy), with periodic boundary conditions (see Suppl. Mat. D \cite{Supplement}).
The ED results of Fig.~\ref{ED}(c) show that the local magnetization is distributed primarily over five spins surrounding 
the defect. A single spin at Cu$_{\mathrm A}$ with a large positive $\langle S_{c}^{\mathrm A} \rangle \sim 0.18-0.30$, 
two spins at Cu$_{\mathrm C}$ with 
also a large postive $\langle S_{c}^{\mathrm C} \rangle \sim 0.18-0.21$, and two spins at Cu$_{\mathrm B}$ 
with a large negative
$\langle S_{c}^{\mathrm B} \rangle \sim -0.1$ add up approximately to the saturated value of 0.5. 
In addition, eight spins at four other sites (Cu$_{\mathrm{D - G}}$) carry a small and oscillating magnetization 
less than 0.1 in absolute value. The local magnetization is much smaller for the remaining sites ($\sim 10^{-3}$)  
and cannot be determined accurately for the cluster sizes of our calculation. 
Interestingly, there is a strong dependence on $\alpha$. First of all,
the polaronic structure collapses very rapidly when $\alpha$ exceeds 0.68, where the pure system undergoes  
a first-order transition from the dimer to the plaquette phase \cite{Koga,Laeuchli,Corboz1}. 
Besides, and more remarkably, 
the magnetization of the unpaired site Cu$_{\mathrm{A}}$, $\langle S_{c}^{\mathrm A} \rangle$, 
strongly depends on $\alpha$. It decreases steeply with $\alpha$ and becomes smaller than 
$\langle S_{c}^{\mathrm C} \rangle$ at $\alpha \sim 0.66$, an observation that will turn crucial 
for the analysis of the experimental spectrum.

To make contact between the local magnetization at Cu sites and the boron spectrum, we note that the
internal field $B_{\mathrm{int}}$ at a given boron site is given by the sum of contributions from neighboring Cu sites
\begin{equation}
B_{\rm int}^i = \sum_{j} A_{ij} \langle S_{c}^{j} \rangle .
\end{equation}
Here $A_{ij}$ is the hyperfine coupling constant from the $i$-th boron site to the $j$-th Cu site. It is the sum of the dipolar 
and transferred hyperfine couplings, $A_{ij} =  D_{ij} + T_{ij}$, and depends on the relative position between the 
boron and Cu sites. The dominant couplings are illustrated in Fig.~\ref{ED}(a) and summarized in Table I.
The transferred hyperfine couplings are short-ranged and limited to the nearest and next-nearest neighbors in the same layer, 
$T_{1}$ and $T_{2}$. They satisfy the condition $T_{1} + 2T_{2} = -0.431$~T 
imposed by the NMR shift data in undoped  SrCu$_2$(BO$_3$)$_2$ \cite{Kodama1}, 
leaving only one adjustable parameter, say $T_{1}$. The analysis of NMR spectra in the 
magnetization plateau phases has led to the estimation 
$-0.71 < T_{1} < -0.53$~T \cite{Takigawa2}. 
The dipolar couplings can be calculated from the crystal parameters.
In addition to the nearest and next neareast neighbors in the same layer, two neighbors on the adjacent
layers have significant dipolar couplings with different values $D_3>D_4$ because of the buckling of the layers. 
Looking at Table I, we can anticipate that the boron sites close to the impurity
both in the layer of the impurity and in the two adjacent layers will have internal fields large enough
to give rise to additional peaks. 

The absolute value of $A_{j}$ is by far the largest for the nearest
neighbor ($j$ = 1). The value of $B_{\mathrm{int}}$ for the B sites in the layer of the impurity should, therefore, 
be primarily determined by $\langle S_{c} \rangle$ of the nearest neighbor Cu site. We then conclude that B$_{1}$ 
and B$_{2}$, which show large negative $B_{\mathrm{int}}$ ($\sim -0.14$~T, see Fig.~\ref{spectrum}), must 
correspond to the boron sites next to either Cu$_{\mathrm{A}}$ or Cu$_{\mathrm{C}}$ in Fig.~\ref{polaron}. 
Likewise, B$_{15}$, with its large positive $B_{\mathrm{int}}$ ($\sim 0.07$~T), 
should be next to Cu$_{\mathrm{B}}$. The values of $\langle S_{c}^{\mathrm{A}} \rangle$ and 
$\langle S_{c}^{\mathrm{C}} \rangle$ can be estimated approximately as $B_{\mathrm{int}}/A_{1} \sim 0.2$, 
which is significantly smaller than the saturated value of 0.5.  
Thus the distribution of $B_{\mathrm{int}}$ provides a direct experimental proof 
for the polaronic spin structure near defects. 

Interestingly, the integrated intensity of the low frequency satellite line of B$_{1}$ at 
87.12~MHz is twice as large as that of B$_{2}$ at 87.24~MHz. Since each Zn impurity 
creates one Cu$_{\mathrm{A}}$ and two Cu$_{\mathrm{C}}$ sites, B$_{1}$ (B$_{2}$) must be 
assigned to boron sites next to Cu$_{\mathrm{C}}$ (Cu$_{\mathrm{A}}$). The larger value of $|B_{\mathrm{int}}|$ at 
B$_{1}$ then leads us to conclude that $\langle S_{c}^{\mathrm{A}} \rangle < \langle S_{c}^{\mathrm{C}} \rangle$. 
Fig.~\ref{ED}(c) shows that this condition is met only in a very narrow range of 
$\alpha$ between 0.655 and 0.68. 

\begin{table}
\caption{\label{tab:table1}Hyperfine coupling constants in Tesla.} \begin{ruledtabular} \begin{tabular}{cccc} $j$&$T_{j}$&$D_{j}$&$A_{j}$\\ \hline
1 & $-0.711 \sim  -0.531$ & $-0.161$ & $-0.872 \sim  -0.692$\\
2 & $0.05 \sim  0.14$ & $-0.075$ & $-0.025 \sim  0.065$\\
3 & 0 & 0.103 & 0.103\\
4 & 0 & 0.065 & 0.065\\
\end{tabular}
\end{ruledtabular}
\end{table}

\begin{figure}[t]
\includegraphics[width=0.9\linewidth]{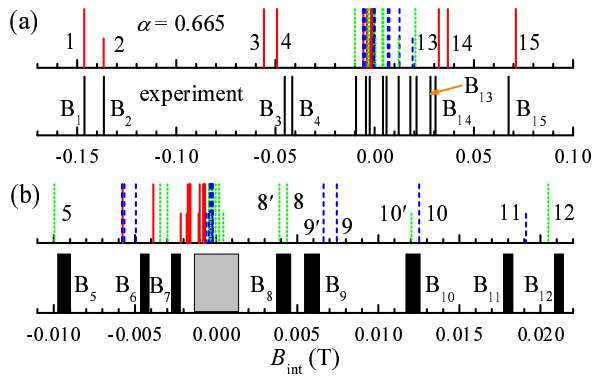}
\caption{(color online) (a) $B_{\mathrm{int}}$ at the B sites 
calculated by using the spin density distribution with $\alpha $ = 0.665 for 36 sites (upper panel) and $B_{\mathrm{int}}$ 
deduced from experiment (lower panel). The data near $B_{\mathrm{int}}$ = 0 T are expanded in Fig.~4(b). The thickness
of the lines in the lower panel indicates the half width at the half maximum of the corresponding peaks of the NMR 
spectrum. In the calculated histograms, red solid lines correspond to B sites in the layer of the Zn impurity, green dotted 
(blue dashed) lines to B in the adjacent layer above (below) the impurity, and the height is proportional to the number of 
B sites having exactly the same internal field (1 or 2).}
\label{histogram}
\end{figure}

Thanks to this assignment, we are now in a position to fix $\alpha$ and $T_1$ by fitting the experimental value of 
$B_{\mathrm{int}}$ at the B$_1$ and B$_2$ sites using the 36-site cluster results (interpolated between 
 $\alpha=0.66$ and $0.67$). This leads to $\alpha=0.665$ and $T_1=-0.563$~T ($A_1=-0.724$, $A_2=-0.009$~T), 
compatible with the values in Table 1. The full theoretical histogram of $B_{\mathrm{int}}$ deduced from Eq.~(2) 
is plotted in the upper panels of Fig.~\ref{histogram}(a) and (b). The isolated red lines in Fig.~\ref{histogram}(a) 
represent $B_{\mathrm{int}}$ at the boron sites in the same layer as the impurity. Each of them is nearest  
to one of the seven Cu sites (Cu$_{\mathrm{A-G}}$) carrying appreciable magnetization. 
The overall agreement between the ED results and experiment is very good, leading to the assignment of
the lines B$_3$, B$_4$, B$_{13}$, B$_{14}$, and B$_{15}$ (see Fig.~\ref{polaron}). 

Since other boron sites in the layer of the impurity have much smaller internal fields, we now turn to 
the neighboring layers. They have smaller values of $B_{\mathrm{int}}$ coming from the interlayer 
dipolar couplings $D_{3}$ or $D_{4}$ as shown in the upper panel of Fig.~\ref{histogram}(b).  
Again, the agreement with the experimental results is very good.
Let us focus on the experimental lines B$_{11}$ and B$_{12}$. 
Since B$_{12}$ is twice as intense as B$_{11}$ (see Fig.~\ref{spectrum}), we must assign B$_{12}$ to the neighbors of 
Cu$_{\mathrm{C}}$ in the layer above, and B$_{11}$ to the neighbor of the Cu$_ {\mathrm{A}}$ in the layer below. 
Since both couplings are given by $D_3$, the larger $B_{\mathrm{int}}$ at B$_{12}$ than B$_{11}$ provides an 
independent confirmation that $\langle S_{c}^{\mathrm{A}} \rangle < \langle S_{c}^{\mathrm{C}} \rangle$. 
With its strongly negative $B_{\mathrm{int}}$, the line $B_5$ must be attributed to the neighbors of Cu$_{\mathrm{B}}$ 
in the layer above. Discussion on the other lines is given in the Suppl. Mat. E \cite{Supplement}. 

So far we have assumed that the hyperfine couplings are not influenced by Zn-doping. However, the small 
difference in the ionic radii between Cu$^{2+}$ and Zn$^{2+}$ (about 5\% \cite{IonicRadii}) could produce non-uniform 
chemical pressure effects, which may result into a local lattice distortion and a modification of the hyperfine couplings.
To estimate such effects, the quadrupole splitting $\nu _Q$ is a useful probe since it is 
sensitive to changes in local structure and charge density.
The inset (b) of Fig.~3 shows the values of $\nu _Q$ for all the observed 
B NMR lines. Remarkably, most sites have exactly the same value $\nu _Q$ = 1.25~MHz as in undoped
SrCu$_2$(BO$_3$)$_2$ (solid line). Only the lines B$_{11}$ and B$_{15}$
show minor deviations of about 0.02~MHz, indicating that the effects of lattice distortion are small 
and limited to the immediate vicinity of the Zn impurities. Note that $\nu _Q$ at the boron sites close to Cu$_{\mathrm{A}}$ 
and Cu$_{\mathrm{C}}$ 
is unchanged, an indication that the hyperfine couplings are likely to remain the same. 
Furthermore, the dipolar coupling, which varies slowly with distance as $1/r^3$, should not be affected 
by a small lattice distortion. Therefore, our conclusion 
$\langle S_{c}^{\mathrm{A}} \rangle < \langle S_{c}^{\mathrm{C}} \rangle$ should remain valid even allowing for a local
distortion around the impurity. 

Finally, let us compare the values of $\alpha$ reported so far from various measurements. 
The analysis of susceptibility and specific heat data of the undoped material by the Shastry-Sutherland model 
with an interlayer coupling has led to the best value $\alpha$ = 0.635 \cite{Miyahara3}, while
the recent determination of the width of the 1/2 plateau in very high 
magnetic fields up to 118~T \cite{Matsuda} led to $\alpha \simeq 0.63$. 
These values are smaller than our estimate $\alpha$ = 0.665 necessary to account for the internal structure of the polaron.
After discarding other possibilities such as Dzyaloshinsky-Moriya (we checked with ED for the Zn doped system that neither the 
intradimer nor the interdimer Dzyaloshinsky-Moriya coupling was able to account for the discrepancy), we came to the conclusion that
the most likely explanation is that the ratio $\alpha$ increases near Zn due to the local
chemical pressure induced by the larger ionic radius of Zn$^{2+}$ as compared to Cu$^{2+}$. Indeed, a similar
effect has already been observed in undoped samples under hydrostatic pressure \cite{ronnow}.
To actually demonstrate that this local modification could explain the discrepancy, we have examined a simple 
model in which the Cu-Cu bond closest to Zn. i.e. the 
Cu$_{\mathrm{A}}$-Cu$_{\mathrm{B}}$ bond $J_{\mathrm{imp}}$ is allowed to change from the bulk $J^{\prime}$ 
(see Suppl. Mat. F \cite{Supplement}). We found that the polaronic structure derived from NMR is actually compatible with 
$\alpha\simeq 0.65$ if $J_{\mathrm{imp}}$ is allowed to take larger values in the range $0.72-0.77$. This value
of $\alpha$ is already significantly lower than the estimate 0.665 for the uniform system, and it sounds plausible
that this value can be further lowered if one allows for additional modifications of the coupling constants. It would be interesting to
investigate this possibility further with the help of ab-initio investigations of the local exchange couplings of Zn doped.
This however goes far beyond the scope of the present paper.

We acknowledge useful discussions with C. Berthier and M. Horvati\'c. 
The work was supported by Grant-in-Aids for 
JSPS KAKENHI (B) (No. 21340093), the MEXT-GCOE program, and the Swiss National Foundation. Numerical simulations were performed at CALMIP and GENCI.

\section*{SUPPLEMENTAL MATERIALS}

\subsection{A. Magnetization due to impurity-induced free spins}

\begin{figure}[t]
\includegraphics[width=0.8\linewidth]{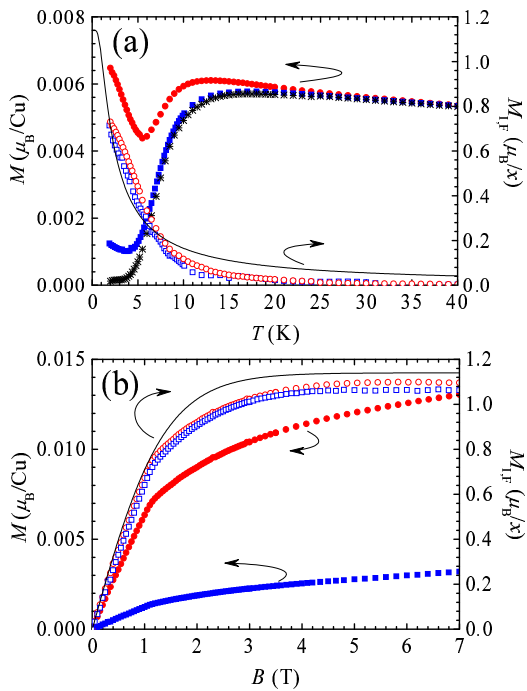}
\caption{(a) Temperature dependences of $M$ at 1 T. (b) Magnetic field dependences of $M$ at 2.0 K. 
The red solid circles, blue solid squares, and asterisks indicate $M$ for $x$ = 0.0174, 0.0032, and 0, respectively. 
The data of $M$ for $x$ = 0 is obtained from Ref.~[26]. 
The red open circles and blue open squares indicate the contribution $M_{\mathrm{I}}$ from impurity-induced spins 
for $x$ = 0.0174 and 0.0032, respectively. The solid line indicates the magnetization of a free spin $M_{\mathrm{F}}$.}
\label{magnetization}
\end{figure}

Figure~\ref{magnetization}(a) shows the temperature dependence of the magnetization $M$ 
of SrCu$_{2-x}$Zn$_x$(BO$_3$)$_2$ at the field of 1~T. The increase of $M$ for $x$ = 0.0174 and 0.0032 at low temperatures
should be ascribed to the impurity-induced unpaired spins.  By subtracting the magnetization for $x$ = 0 from these
data and normalizing by $x$, we obtain the contributions from the impurity-induced spins $M_{\mathrm{I}}$, which 
are plotted in Fig.~\ref{magnetization}(a). The values of $M_{\mathrm{I}}$ are nearly identical for $x$ = 0.0174 and 0.0032 
in the whole temperature range, indicating that Zn$^{2+}$ ions effectively replace the Cu sites. 
The increase of $M_{\mathrm{I}}$ below 10 K is described reasonably well by a free spin model
$M_{\mathrm{F}} = gS\mu _BB_S(X)$ with $S$ = 1/2, where $B_S(X)$ ($X = gS\mu _BB/k_BT$) is the Brillouin function. 
Above 10~K, on the other hand, $M_{\mathrm{I}}$ is much smaller than $M_{\mathrm{F}}$,
indicating that Zn impurites can no longer generate free spins because of interaction between unpaired Cu 
spins and thermally excited triplets.  

\begin{figure}[t]
\includegraphics[width=0.7\linewidth]{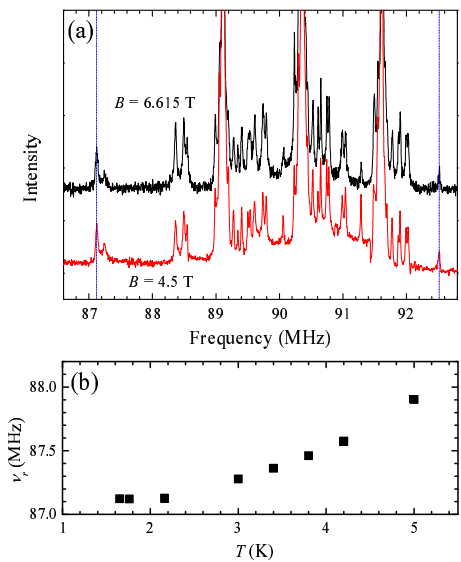}
\caption{(a) $^{11}$B NMR spectra of SrCu$_{2-x}$Zn$_x$(BO$_3$)$_2$ with $x$ = 0.0174 at 1.6~K for 
two different field values. The frequency for the spectrum at 4.5~T is shifted by 28.87~MHz to account for the 
difference in the nuclear Zeeman frequency. The position of the lines then match perfectly, indicating that the 
distribution of the spin moments remains the same for the two field values, hence that the impurity induced 
moments are saturated. (b) Temperature dependence of the peak frequency of the low frequency quadrupole satellite line 
of B$_1$ sites obtained at 6.615~T.  It remains constant below 2.2~K, again indicating saturation of the 
impurity-induced moments.}
\label{saturation}
\end{figure}

Figure~\ref{magnetization}(b) shows the magnetic field dependence of $M$ at 2 K. Although the magnetization of a free 
spin $M_{\mathrm{F}}$ saturates completely above 4~T as indicated by the solid line, $M$ of  
SrCu$_{2-x}$Zn$_x$(BO$_3$)$_2$ keeps increasing almost linearly with $B$ at high fields. The NMR spectra shown in 
Fig.~\ref{saturation}(a), on the other hand, indicate that the internal fields at boron sites stay exactly the same 
between 4.5 and 6.615~T, a clear indication of the saturation of the spin moments. The temperature dependence of the 
resonance frequency shown in Fig.~\ref{saturation}(b) provides further support for the saturation of spin moments. 
Thus the linear increase of $M$ at high fields cannot be attributed to spin moments. It may be associated with 
orbital (van Vleck) magnetism, even though we do not understand the mechanism for such a behavior. 
By subtracting the $B$-linear component at high fields from $M$, we obtain the contributions 
of the unpaired spins $M_{\mathrm{I}}$, which are normalized by $x$ and are plotted in Fig.~\ref{magnetization}(b). 
There is almost no difference between $M_{\mathrm{I}}$ for $x$ = 0.0174 and 0.0032. They also 
agree reasonably well with the free spin behavior $M_{\mathrm{F}}$.  

\subsection{B. Dynamics of the impurity induced spins near saturation}

When the impurity-induced moments become saturated as the temperature is decreased, we expect the thermal fluctuations 
to be gradually depressed and to slow down. Such a process has been indeed confirmed by measurements of 
the nuclear spin-lattice relaxation rate 1/$T_1$ and of the spin-echo decay rate 1/$T_2$. A standard inversion recovery 
method was used for the 1/$T_1$ measurement. To determine 1/$T_2$, the spin echo intensity $I(\tau)$ as a function 
of the time $\tau$ between the two rf-pulses was fit to the exponential function $I(\tau ) = C_0\mathrm{exp}(-2\tau /T_2)$. 

\begin{figure}[b]
\includegraphics[width=0.7\linewidth]{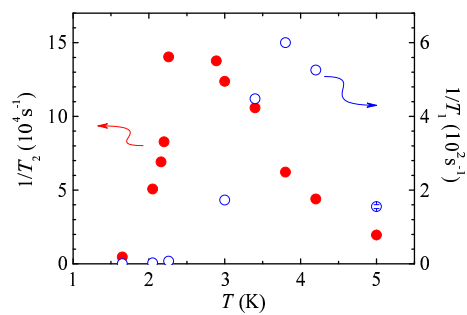}
\caption{Temperature dependence of 1/$T_1$ (blue open circles) and 1/$T_2$ (red solid circles) 
measured on the low frequency quadrupole satellite line of the B$_1$ site at 6.615 T.}
\label{dynamics}
\end{figure}

Figure~\ref{dynamics} shows the temperature dependences of 1/$T_1$ and 1/$T_2$ measured on the low frequency 
quadrupole satellite line of the B$_1$ site at 6.615 T. Both 1/$T_1$ and 1/$T_2$ exhibit a peak but at different 
temperatures. The peak in 1/$T_1$ occurs at 3.8~K while the peak in 1/$T_2$ appears near 2.5~K. In fact, $T_2$ becomes 
too short near 2.5~K to be able to observe NMR signal. At lower temperatures, both 1/$T_1$ and 1/$T_2$ show steep 
decrease, consistent with the excitation gap for impurity induced spins in a magnetic field. 

Let us first discuss 1/$T_1$, which is related to the time correlation function of the local field by the 
standard formula, $1/T_1 \propto G(\omega_{\mathrm N})$, where $G(\omega) =  \int \langle h(0)h(t) \rangle 
\mathrm{exp}(i\omega t)dt$ and $\omega_{\mathrm N}$ is the NMR frequency.  A simple expression for 
the correlation function parametrized by the mean square amplitude of the fluctuation $\langle h^2 \rangle$  
and the correlation time $\tau _c$, $\langle h(0)h(t) \rangle = \langle h^2 \rangle \exp(-t/\tau_c)$ leads to  
the following simple result,
\begin{equation}
1/T_1 \propto <h^2> \frac{\tau_c}{1+(\omega_{\mathrm N} \tau_c)^2}.
\end{equation} 
The saturation of the impurity-induced moments is expected to proceed with the depression of $\langle h^2 \rangle$
and the growth of $\tau _c$. The peak in 1/$T_1$ can be reproduced only by the latter process. Indeed 1/$T_1$ exhibits
a peak when $\tau_c = 1/\omega_{\mathrm N} \sim 1.8 \times 10^{-9}$~s, a well known result in the context of motional
narrowing in classical NMR [29].  

The slowing down of the spin fluctuation can also cause a peak in 1/$T_2$, with different criteria however. 
When the fluctuating local field slows down, the fastest spin-echo decay is achieved when $1/\tau_c$ becomes 
comparable to $\gamma \sqrt{\langle h^2 \rangle}$ and the peak value of 1/$T_2$ has the same orders 
of magnitude as $1/\tau_c$ at the peak temperature [30, 31]. 
Although we were not able to 
determine the peak value of 1/$T_2$ due to loss of NMR signal, it should be of the order of $10^6$~s$^{-1}$,
judging from the data in Fig.~\ref{dynamics}, indicating that $\tau_c \sim 10^{-6}$~s at 2.5 K, which is much 
longer than the value estimated at the peak temperature of 1/$T_1$  (3.8~K). Thus, different peak 
temperatures of 1/$T_1$ and 1/$T_2$ provide evidence for a rapid but gradual slowing down of the 
fluctuation of the impurity-induced spins, which  precedes the complete saturation at lower temperatures.

\subsection{C. Comparison of the NMR spectra for different Zn concentration}

\begin{figure}[b]
\includegraphics[width=0.8\linewidth]{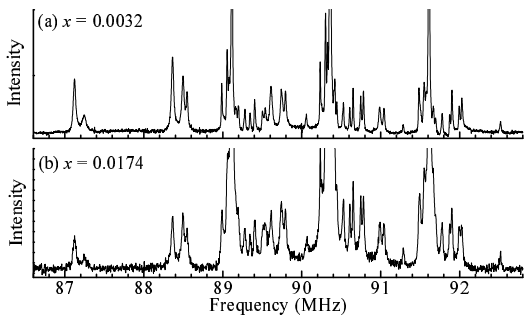}
\caption{$^{11}$B NMR spectra of SrCu$_{2-x}$Zn$_x$(BO$_3$)$_2$ with $x$ = 0.0032 (a) and 0.0174 (b) 
at 1.6 K and 6.615 T.}
\label{xdep}
\end{figure}

Figure~\ref{xdep} shows the $^{11}$B NMR spectra for $x$ = 0.0032 and 0.0174 
at 1.6 K and 6.615 T. Both samples show nearly identical spectra. All the resolved peaks are exactly at the 
same frequencies, indicating that the average distance between Zn$^{2+}$ ions is sufficiently large  
and effects of interaction between impurity-induced spins can be neglected.  
However, the widths  of the individual lines for $x$ = 0.0174 are slightly larger than those for $x$ = 0.0032, 
which is likely due to the difference in the distribution of the demagnetizing fields inside the crystals.

\subsection{D. Numerical Methods}

\begin{figure}[t]
\includegraphics[width=0.6\linewidth]{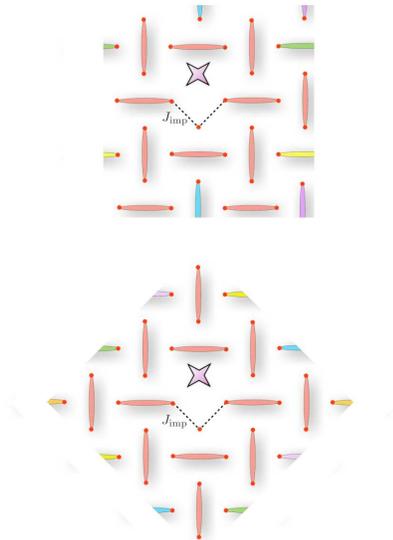}
\caption{Clusters used for the simulations with $32$ sites (top) and 36 sites (bottom) and one impurity represented as a cross. Colored dimers correspond to strong $J$ bonds while $J'$ bonds are not shown for clarity (except the 
$J_\mathrm{ imp}$ bonds introduced in the main text). Periodic boundary conditions are used to minimize finite-size effects, so that colored dimers crossing the boundaries can be easily identified with different colors. }
\label{clusters}
\end{figure}

Exact Diagonalization (ED) have been performed using the standard Lanczos algorithm for 
the $S=1/2$ Heisenberg model on Shastry-Sutherland lattices ($N=32$ and 36 sites, 
see Fig.~\ref{clusters}) having a single impurity. Periodic boundary conditions have been 
used to minimize finite-size effects. The presence of an impurity forbids to use any translation 
symmetry, but we have still made use of the remaining reflection symmetry to reduce the size of 
the Hilbert space. After computing the ground-state, one can easily extract the average magnetization 
on each site. Note that since the polaron has a large but finite extension beyond which the magnetization
becomes very small, one can expect a priori the finite-size effects to be small. This is confirmed by a comparison
of the results obtained on 32-site and 36-site clusters, for which the magnetizations are nearly identical,
a conclusion further confirmed by calculations on larger clusters up to 64 sites with infinite Product Entangled
Pair State simulations [P. Corboz, private communication]. We are thus  confident that 
the local magnetizations reported in this work are close to their thermodynamical values (see Fig.~3c of the main text). 

\subsection{E. Intensity and line shape of some specific lines.}

\begin{figure}[t]
\includegraphics[width=0.6\linewidth]{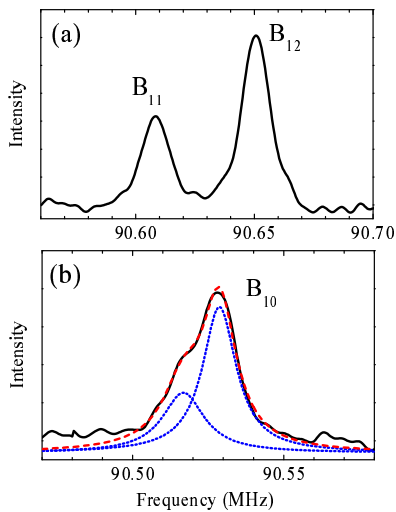}
\caption{Enlarged NMR spectra for the center lines of (a) B$_{11}$ and B$_{12}$ and (b) B$_{10}$ lines..}
\label{intensity}
\end{figure}

One of our main conclusion that $\langle S_{c}^{\mathrm{A}} \rangle < \langle S_{c}^{\mathrm{C}} \rangle$ is based on 
the intensity ratio of the B$_1$ and B$_2$ lines, which are assigned to the B sites close to Cu$_{\mathrm{C}}$ and
Cu$_{\mathrm{A}}$ on the layer of the Zn impurity. Equally important is the intensity ratio of of the 
B$_{11}$ and B$_{12}$ lines assigned to the B sites on the adjacent layers coupled to Cu$_{\mathrm{C}}$ 
and Cu$_{\mathrm{A}}$ with the same dipolar coupling $D_3$. The spectrum of the center line of these sites are 
displayed in Fig.~\ref{intensity}(a) with an enlarged scale.  We can clearly see that the integrated intensity of the 
B$_{12}$ line is twice as large as the B$_{11}$ line. Since Cu$_{\mathrm{C}}$ is twice as abundant as Cu$_{\mathrm{A}}$, 
the B$_{12}$ (B$_{11}$) line must be assigned to the neighbor of Cu$_{\mathrm{C}}$ (Cu$_{\mathrm{A}}$). The fact that 
$B_{\mathrm{int}}$ is larger for the B$_{12}$ line than for the B$_{11}$ line then leads to the conclusion that 
$\langle S_{c}^{\mathrm{A}} \rangle < \langle S_{c}^{\mathrm{C}} \rangle$.

The ED calculation shows that there is another pair of lines (10 and 10$^{\prime}$) assigned to the B sites on the adjacent
layers coupled to Cu$_{\mathrm{C}}$ and Cu$_{\mathrm{A}}$ with the smaller dipolar coupling $D_4$ (see the 
upper panel of Fig.~4b of the main text). Although the experimentally observed B$_{10}$ line shows only a single peak, 
the spectral shape can be actually fit well to a sum of two Lorentzians with the intensity ratio of 2 to 1 as indicated 
in Fig.~\ref{intensity}(b).  The fact that one of these lines with larger $B_{\mathrm{int}}$ has the double intensity is again 
consistent with the conclusion $\langle S_{c}^{\mathrm{A}} \rangle < \langle S_{c}^{\mathrm{C}} \rangle$.

There is a puzzle though. Since the two pairs of lines are coupled to the same pair of Cu sites (Cu$_{\mathrm{C}}$ and
Cu$_{\mathrm{A}}$) with different dipolar couplings, their separation should be equal to the ratio of the dipolar couplings
$D_3/D_4=1.58$. However, this ratio in the experimental spectrum is much larger (nearly 3). This suggests that the
internal fields at the B sites in the adjacent layers are not entirely due to the in-plane Cu moments and that small 
contributions from induced moments on Cu sites in the adjacent layers due to interlayer exchange may be relevant. 
To explore such effects is beyond the scope of this paper however. 

For other experimental lines B$_{6}$ -  B$_{9}$, the resolution is not sufficient to achieve a one to one 
correspondence with the calculated lines. However, the assignment of lines B$_8$ and B$_9$ to the pairs of 
theoretical lines (8, 8$^{\prime}$) and (9, 9$^{\prime}$) is plausible since they are well 
separated from the others. Finally, the lines B$_6$ and B$_7$ develop on top of the broad tail of the main line, consistent with 
the numerous theoretical lines on the negative side of the main line, even though specific assignment is not possible. 

\subsection{F. Possible local change of exchange interaction due to lattice distortion}

\begin{figure}[t]
\includegraphics[width=0.7\linewidth]{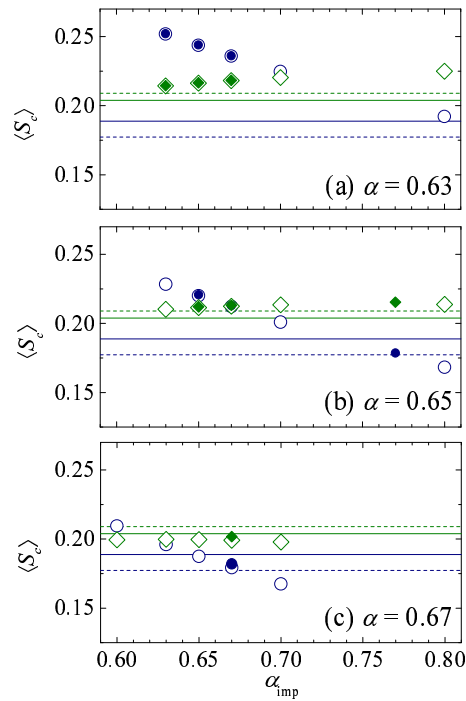}
\caption{Local magnetization at Cu$_{\mathrm{A}}$ ($\langle S_{c}^{\mathrm{A}} \rangle$, circles) and
Cu$_{\mathrm{C}}$ ($\langle S_{c}^{\mathrm{C}} \rangle$, squares) calculated by exact diagonalization for a cluster 
with 32 (open symbols) or 36 (filled symbols) sites are plotted against 
$\alpha_{\mathrm{imp}}=J_{\mathrm{imp}}/J$ for three values of bulk $\alpha=J^{\prime}/J$.
The horizontal lines indicate the values of $\langle S_{c}^{\mathrm{A}} \rangle$ and ($\langle S_{c}^{\mathrm{C}} \rangle$
determined from the values of $B_{\mathrm{int}}$ at the B sites on the same layers as the Zn impurity (the solid lines) or 
on the neighboring layers (the dashed lines).}
\label{distortion}
\end{figure}

Here we employ a simple model to account for a possible change of exchange interaction due to local lattice 
distortion near Zn impurities and examine to what extent our best choice of $\alpha=0.665$ is influenced by 
such effects. Since the distribution of $\nu_Q$ shown in Fig.~3(b) of the main text suggests that the 
lattice distortion is confined in the immediate vicinity of Zn, the simplest model consists in assuming that only the 
exchange interaction for the Cu-Cu bonds closest to the Zn impurity, i.e. the exchange between Cu$_{\mathrm{A}}$ and 
Cu$_{\mathrm{B}}$, changes from the bulk $J^{\prime}$ to $J_{\mathrm{imp}}$ (see Fig.~\ref{clusters}). 
   
We have calculated the moments at Cu$_{\mathrm{A}}$ and Cu$_{\mathrm{C}}$ sites by exact diagonalization 
using a cluster of 32 or 36 sites as a function of $\alpha_{\mathrm{imp}}=J_{\mathrm{imp}}/J$ 
for three values of the bulk ratio $\alpha=J/J^{\prime}$, 0.63, 0.65, and 0.67. 
The results plotted in Fig.~\ref{distortion} indicate that $\langle S_{c}^{\mathrm{A}} \rangle$
decreases rapidly with $\alpha_{\mathrm{imp}}$. However,  $\langle S_{c}^{\mathrm{C}} \rangle$ stays nearly constant or 
even increases slightly with $\alpha_{\mathrm{imp}}$. Therefore, the condition 
$\langle S_{c}^{\mathrm{A}} \rangle < \langle S_{c}^{\mathrm{C}} \rangle$ can be met for $\alpha$ smaller than 0.66 if 
we allow for a large value of $\alpha_{\mathrm{imp}}$ as demonstrated in Fig.~\ref{distortion}. 

To make a quantitative comparison with the experimental data, we extracted the local magnetization from the observed 
values of $B_{\mathrm{int}}$. First, we used the values of $B_{\mathrm{int}}$ at seven B sites on the layer of the impurity 
(B$_{1-4}$ and B$_{13-15}$ in Fig.~4b of the main text), each of which is the nearest neighbor to one of the major Cu 
sites (Cu$_{\mathrm{A-G}}$) carrying appreciable moments. The values of $\langle S_{c}^{\mathrm{A-G}} \rangle$ 
can be determined by solving Eq.~(2) after substituting the experimental values of B$_{\mathrm{int}}$ into the left side 
and using  the same values of the hyperfine coupling constants $T_1 = -0.563$~T.  The values of Cu$_{\mathrm{A}}$ and 
Cu$_{\mathrm{C}}$ thus determined are shown by the solid lines in Fig.~\ref{distortion}. Alternatively, we can use the values 
of $B_{\mathrm{int}}$ at the B sites in the adjacent layers (B$_{5}$, B$_{11}$, and B$_{12}$) coupled to the three major Cu 
sites (Cu$_{\mathrm{A-C}}$) by the dipolar coupling $D_3$ to determine $\langle S_{c}^{\mathrm{A-C}} \rangle$. The results 
are displayed by the dashed lines in Fig.~\ref{distortion}. From these plots, we may conclude that the experimental results 
can be reconciled with $\alpha=0.65$ by allowing a rather large modification of exchange coupling near Zn. 
The value $\alpha=0.63$ seems difficult to reconcile with the experimental results within this simple model, but these results
make it plausible that  allowing for additional modifications slightly further away from Zn can further reduce the value of $\alpha$.


\begin{thebibliography}{99}
\bibitem{Alloul} H. Alloul, J. Bobroff, M. Gabay, and P. J. Hirschfeld, Rev. Mod. Phys. \textbf{81}, 45 (2009).
\bibitem{Hagiwara} M. Hagiwara, K. Katsumata, I. Affleck, B. I. Halperin, and J. P. Renard, Phys. Rev. Lett. \textbf{65}, 3181 (1990).
\bibitem{Glarum} S. H. Glarum, S. Geschwind, K. M. Lee, M. L. Kaplan, and J. Michel, Phys. Rev. Lett. \textbf{67}, 1614 (1991).
\bibitem{Yoshida} M. Yoshida, K. Shiraki, S. Okubo, H. Ohta, T. Ito, H. Takagi, M. Kaburagi, and Y. Ajiro, Phys. Rev. Lett. \textbf{95}, 117202 (2005).
\bibitem{Tedoldi} F. Tedoldi, R. Santachiara, and M. Horvati\'{c}, Phys. Rev. Lett. \textbf{83}, 412 (1999).
\bibitem{Das} J. Das, A. V. Mahajan, J. Bobroff, H. Alloul, F. Alet, and E. S. S\o rensen, Phys. Rev. B \textbf{69}, 144404 (2004).
\bibitem{Takigawa} M. Takigawa, N. Motoyama, H. Eisaki, and S. Uchida, Phys. Rev. B \textbf{55}, 14129 (1997).
\bibitem{Kudo}K. Kudo, T. Noji, Y. Koike, T. Nishizaki, and N. Kobayashi, J. Phys. Soc. Jpn. \textbf{73}, 3497 (2004).
\bibitem{Shawish} S. El Shawish and J. Bon\v{c}a, Phys. Rev. B \textbf{74}, 174420 (2006).
\bibitem{Haravifard} S. Haravifard, S. R. Dunsiger, S. El Shawish, B. D. Gaulin, H. A. Dabkowska, M. T. F. Telling, T. G. Perring, and J. Bon\v{c}a, Phys. Rev. Lett. \textbf{97}, 247206 (2006).
\bibitem{Capponi} S. Capponi, D. Poilblanc, and F. Mila, Phys. Rev. B \textbf{80}, 094407 (2009).
\bibitem{Shastry} B. S. Shastry and B. Sutherland, Physica B+C \textbf{108}, 1069 (1981).
\bibitem{Kageyama1} H. Kageyama, K. Yoshimura, R. Stern, N. V. Mushnikov, K. Onizuka, M. Kato, K. Kosuge, C. P. Slichter, T. Goto, and Y. Ueda, Phys. Rev. Lett. \textbf{82}, 3168 (1999).
\bibitem{Kodama1} K. Kodama, J. Yamazaki, M. Takigawa, H. Kageyama, K. Onizuka, and Y. Ueda, J. Phys.: Condens. Matter \textbf{14}, L319 (2002).
\bibitem{Miyahara1} S. Miyahara and K. Ueda, Phys. Rev. Lett. \textbf{82}, 3701 (1999).
\bibitem{Koga} A. Koga and N. Kawakami, Phys. Rev. Lett. \textbf{84}, 4461 (2000).
\bibitem{Corboz1} P. Corboz and F. Mila, Phys. Rev. B \textbf{87}, 115144 (2013).
\bibitem{Onizuka}K. Onizuka, H. Kageyama, Y. Narumi, K. Kindo, Y. Ueda, and T. Goto,  J. Phys. Soc. Jpn. \textbf{69}, 1016 (2000).
\bibitem{Kodama2} K. Kodama, M. Takigawa, M. Horvati\'{c}, C. Berthier, H. Kageyama, Y. Ueda, S. Miyahara, F. Becca, and F. Mila, Science \textbf{298}, 395 (2002). 
\bibitem{Sebastian} S. E. Sebastian, N. Harrison, P. Sengupta, C. D. Batista, S. Francoual, E. Palm, T. Murphy, N. Marcano, H. A. Dabkowska,
and B. D. Gaulin, Proc. Natl. Acad. Sci. USA \textbf{105}, 20157 (2008).
\bibitem{Jaime} M. Jaime, R. Daou, S. A. Crooker, F. Weickert, A. Uchida, A. E. Feiguin, C. D. Batista,
H. A. Dabkowska, and B. D. Gaulin, Proc. Natl. Acad. Sci. {\textbf 109}, 12404 (2012). 
\bibitem{Takigawa2} M. Takigawa, M. Horvati\'{c}, T. Waki, S. Kramer, C. Berthier, F. Levy-Bertrand, I. Sheikin, H. Kageyama, Y. Ueda, and F. Mila, Phys. Rev. Lett. \textbf{110}, 067210 (2013).
\bibitem{Matsuda} Y. H. Matsuda, N. Abe, S. Takeyama, H. Kageyama, P. Corboz, A. Honecker, S. R. Manmana, G. R. Foltin, K. P. Schmidt, 
and F. Mila, Phys. Rev. Lett. \textbf{111}, 137204 (2013). 
\bibitem{Miyahara3} For an early review, see S. Miyahara and K. Ueda, J. Phys.: Condens. Matter \textbf{15}, R327 (2003).
\bibitem{Takigawa4} For a recent review, see M. Takigawa and F. Mila, \textit{Introduction to Frustrated Magnetism}, edited by C. Lacroix, P. Mendels, and F. Mila (Springer, New York, 2011), p. 241.
\bibitem{Kageyama2} H. Kageyama, K. Onizuka, T. Yamauchi and Y. Ueda, J. Cryst. Growth \textbf{206}, 65 (1999).
\bibitem{Kudo2} K. Kudo, T. Noji, Y. Koike, T. Nishizaki and N. Kobayashi, J. Phys. Soc. Jpn. \textbf{70}, 1448 (2001).
\bibitem{Supplement} See the supplemental material at *********, which includes Refs. \cite{Kageyama2,Abragam,Takigawa86,Recchia}, for details about the experiments and the theory. 
\bibitem{Abragam} A. Abragam, ``{\it The principles of Nuclear Magnetism}'' (Oxford Univ. Press, 1961). 
\bibitem{Takigawa86} M. Takigawa and G. Saito, J. Phys. Soc. Jpn. \textbf{55}, 1233 (1986). 
\bibitem{Recchia} C. H. Recchia, K. Gorny, and C. H. Pennington, Phys. Rev. B \textbf{54}, 4207 (1996). 
\bibitem{Laeuchli} A. L\"auchli, S. Wessel, and M. Sigrist, Phys. Rev. B \textbf{66}, 014401 (2002). 
\bibitem{IonicRadii} R.D. Shannon, Acta Cryst. A \textbf{32}, 751 (1976). 
\bibitem{ronnow} M. E. Zayed, PhD thesis (EPFL, 2010); M.E. Zayed, Ch. R\"uegg, E. Pomjakushina, M. Stingaciu, K. Conder, M. Hanfland, M. Merlini, H.M. Ronnow, Solid State Comm. \textbf{186}, 13 (2014).

\end{thebibliography}
\end{document}